\documentstyle[preprint,aps]{revtex}

\begin{document}


\def \st{\tilde{t}}
\def \mstop#1{\widetilde{M}_{t_{#1}}}
\def \mneut#1{\widetilde{M}_{N_{#1}}}
\def \mLSP{\widetilde{M}_{LSP}}
\def \neut#1{\widetilde{N}_{#1}}
\def \char#1#2{\widetilde{C}_{#1}^{#2}}
\def \mchar#1{\widetilde{M}_{C_{#1}}}
\def \msnu{\widetilde{M}_{\nu}}
\def \mslp#1{\widetilde{M}_{\ell_{#1}}}
\def \snu{\tilde{\nu}}
\def \slp{\tilde{\ell}}
\def \sel#1#2{\tilde{e}_{#1}^{#2}}
\def \ETmiss{{\not\negthinspace\negthinspace{E}_T}}
\def \Emiss{{\not\negthinspace\negthinspace{E}}}
\def \mmiss{\thinspace\thinspace{\not{\negthinspace\negthinspace M}}}
\def \BR{{\cal B}}
\def \heavychar{(i)}
\def \heavystop{(ii)}
\def \degenerate{(iii)}

\def \GeV{{\rm \enspace GeV}}
\def \ts{\thinspace}
\def \beq{\begin{equation}}
\def \eeq{\end{equation}}
\def \beqa{\begin{eqnarray}}
\def \eeqa{\end{eqnarray}}
\def \SM{Standard Model}

\draft
\preprint{
  \parbox{1.4in}{UM--TH--97--08 \\
  [-0.12in] hep-ph/9704450
}  }

\title{Recognizing Superpartners at LEP}
\author{G.L. Kane \cite{GLKemail} and Gregory Mahlon \cite{GDMemail}}
\address{Department of Physics, University of Michigan \\
500 E. University Ave., Ann Arbor, MI  48109 }
\date{April 30, 1997}
\maketitle
\begin{abstract}
There is a class of supersymmetric models which is  
well-motivated by hints of evidence for SUSY and consistent
with all existing data.  
It is important to study the predictions of these models.
They are characterized by 
$\mneut{3} \agt \mchar{1} > \msnu > \mneut{1}$
(where $\neut{i}$ and $\char{i}{}$ are neutralino and 
chargino mass eigenstates), 
$\vert\mu\vert\alt M_1 \alt M_2 \approx M_Z$,
$\mu < 0$, and $\tan\beta$ near 1.  
Their LEP signatures are mostly unusual.
Most produced superpartners are invisible!  A good
signature is two photons plus large missing energy.
There are also excess events at large recoil mass in the
single photon plus nothing channel.
We list the main signatures for charginos,
stops, etc., which are also likely to be unconventional.
This class of models will be definitively tested at LEP194
with 100 pb$^{-1}$ per detector, and almost definitively
tested at LEP184.

\end{abstract}
\pacs{}


\section{Introduction}

How might superpartners be detected if they exist?
They could be found in direct production at colliders as
energy or luminosity is increased and a threshold is crossed
in either.  Observing them of course requires triggering on
such events, and separating the signal from backgrounds.  Or, their
effects could be seen from one-loop contributions.  
Most possible deviations from the \SM, whether particle
production or loop effects, could not be
interpreted as signals of supersymmetry,
since supersymmetric signals are strongly constrained as to which
processes they can contribute to.
For example, SUSY production events should have missing energy 
(assuming $R$-parity conservation), while deviations caused
by loop effects should show up most strongly in 
processes such as $Z\rightarrow b\bar{b}$ and $b\rightarrow s\gamma$
(but {\it not}\ in $b$ quark asymmetries).

In the past couple of years some evidence of such effects
has been reported\cite{SUSYhints}.  Here we do not wish to
describe this evidence or argue that it is, in fact,
evidence for superparticles, but to observe that
it can only be interpreted as such evidence and
be consistent with all existing data if certain 
parameters lie in certain ranges.  Therefore, it is 
appropriate to study the predictions of models with such 
parameters.  We can summarize these
ranges approximately by
\beqa
&& \vert\mu\vert\alt M_1 \alt M_2 \approx M_Z \alt \mslp{L} < \mslp{R}, 
    \quad \mu<0, \cr &&
\qquad\qquad\qquad 1.2 \alt \tan\beta \alt 1.8, \cr &&
\qquad\qquad\qquad \mstop{1} \approx \mstop{R} \sim M_Z.
\label{Inputs}
\eeqa
We focus upon models satisfying the following mass hierarchy,
which results for many combinations of the above inputs:
\beqa
&& \mneut{3} \agt \mchar{1} > \msnu > \mneut{2} > \mneut{1}, \cr &&
\enspace\qquad 45 \alt \mneut{1} \alt 65 \GeV.
\label{MassH}
\eeqa
Here $\mneut{i}$ and $\mchar{i}{}$ are the (unsigned) neutralino
and chargino mass eigenvalues, 
$\mslp{L}$ and $\mslp{R}$ are the left and right charged slepton
masses, $\msnu$ is the sneutrino mass,
$M_1$ and $M_2$ are the $U(1)$ 
and $SU(2)$ gaugino masses at the electroweak scale,
$\mu$ is the coefficient of the $H_U H_D$ term in the
superpotential at the electroweak scale, and $\tan\beta$
is the ratio of the Higgs vevs.  
We also take
the right slepton mass $\mslp{R}$ to be of order 125 GeV,
but this choice only indirectly affects the LEP analysis.
We further impose
$\mneut{2}-\mneut{1} \agt 20 \GeV$, so that we
qualitatively expect events such as 
$\ell^{+}\ell^{-}\gamma\gamma\ETmiss$
at Fermilab\cite{eegamgam}.
The models discussed here do not depend on the light
stop mass indicated in~(\ref{Inputs}) for their interest.
However, the relative size of $\mstop{1}$ and $\mchar{1}$
affects the signatures considerably, so we examine all cases.
We use the word ``models'' to denote different choices of the above
parameters, all constrained to be in the ranges specified
by~(\ref{Inputs}) and~(\ref{MassH}).  Our entire analysis is in terms
of an effective general lagrangian
at the weak scale, including soft supersymmetry
breaking parameters.

The scenario defined by (\ref{Inputs}) is at present a well-motivated
one to study for its implications for signals at
LEP (and Fermilab), and that is the purpose of this
paper.  Further, the detectable signatures for LEP are rather
non-standard.  For example, the largest SUSY signals may
occur as events with two photons and large missing
energy, an excess of single photon events with recoil
mass above about 140 GeV, an excess of events with two soft
charged leptons and large missing energy, etc.
Even though some superpartner production cross sections are 
large (of order 1 picobarn), giving hundreds of produced events,
most are simply invisible!  
The signals that have been most often studied from charginos
and neutralinos are relatively small in this scenario.
The key features which give these models their unique character
are large branching ratios to invisible final states of the
sneutrinos and $\neut{3}$, as well as a significant branching
ratio for the radiative decay of $\neut{2}$.
In the following we
first summarize the masses and decays predicted for
individual superpartners, and then describe the resulting
cross sections and signatures.


\section{Sparticle Properties} \label{Properties}

Table~\ref{Decays} summarizes the properties of the 
relevant superpartners.
The constraints in Eq.~(\ref{Inputs}) imply\cite{longNLSP} that
$\neut{1}$ is dominantly higgsino (the approximately
symmetric combination of $H_U$ and $H_D$), 
$\neut{2}$ is mainly photino, and
$\neut{3}$ is mostly antisymmetric higgsino, but with
a non-negligible zino component. 
The charginos are neither pure wino nor pure charged higgsino
(The $U$ mixing matrix is approximately off-diagonal, while
all of the $V$  mixing matrix elements are roughly the same
magnitude).
All of the results of Table~\ref{Decays} follow straightforwardly.
We have included a row for the stop
since it is motivated
if there are deviations from the  \SM\ in $R_b$ and
$\BR(b\rightarrow s\gamma)$\cite{SUSYhints}, by 
electroweak baryogenesis\cite{Baryogenesis},
and to some observers by aspects of Fermilab data\cite{XtraTops,SUwgt}.
However, the crucial features of the class of models we are studying
do not change if we take the stop mass to be heavier than
indicated in Table~\ref{Decays}.

As described in detail in Ref.~\cite{SandroLong},
an important feature of the SUSY parameters chosen
in~(\ref{Inputs}) is a large value of 
$\BR(\neut{2}\rightarrow\neut{1}\gamma)$.
Consequently, photons will play a significant role in
the signals within this class of models.  The remainder
of the $\neut{2}$ decay rate is to 3-body final states.
Of the possible 3-body decays,
$\neut{2}\rightarrow\neut{1}\ell^{+}\ell^{-}$ has the largest branching
ratio, because of the relatively low slepton masses.

A key result of the mass hierarchy~(\ref{MassH})
concerns the sneutrino decay modes.  The only two-body
modes open are $\snu\rightarrow\neut{1}\nu$ and
$\snu\rightarrow\neut{2}\nu$.  In fact, the decay to
$\neut{1}\nu$ dominates, rendering the sneutrino
almost entirely invisible.
Likewise, since $\neut{3}\rightarrow \snu\nu$
dominates, $\neut{3}$ is also mainly invisible.
Since $e^{+}e^{-} \rightarrow \snu\bar{\snu}$ and
$e^{+}e^{-} \rightarrow \neut{1}\neut{3}$ are among the largest
cross sections, most sparticle production at LEP is
invisible.  When $\snu$ or $\neut{3}$ do have a visible decay,
it contains a single $\gamma$, or possibly 
$\neut{3}\rightarrow\slp_L^{\pm}\ell^{\mp}
\rightarrow\neut{2}\ell^{+}\ell^{-}
\rightarrow\neut{1}\gamma\ell^{+}\ell^{-}$.

The heaviest neutralino, $\neut{4}$, has decays which
are very similar to $\neut{3}$.   The dominant $\neut{4}$ decay mode
is $\snu\nu$. 
Up to a quarter of the total $\neut{4}$ decay rate is to 
$\slp_L\ell$.  So while most $\neut{4}$'s are invisible, 
a significant fraction will produce $\ell^{+}\ell^{-}\gamma\Emiss$.
If the Higgs mass is small enough and the 
$\neut{4}$-$\neut{1}$ mass splitting large enough, the decay
$\neut{4}\rightarrow\neut{1}h^0$ opens up, suggesting the 
possibility
of a significant unconventional source of Higgs bosons at LEP.
However, the
amount of available phase space is small, limiting 
this branching ratio to at most a percent or two, rendering
such prospects dim. 

Charginos and stops decay very differently depending on their
relative mass.   Here we focus upon $\st_1$ and $\char{1}{}$ 
since $\st_2$ and $\char{2}{}$ are likely to be too massive
to be important at LEP in the short term.
There are three interesting possibilities:
\heavychar\ $\mchar{1} > \mstop{1}{+} m_b$,
\heavystop\ $\mstop{1} > \mchar{1}{+} m_b$,
and \degenerate\ $\mstop{1}{+}m_b > \mchar{1} > \mstop{1}{-}m_b$.
In region \heavychar\,
the stop decays exclusively to $c\neut{1}$, as
no other two-body modes are kinematically allowed.  
For the chargino, both $\ell\snu$ and $\st_1 b$ final
states are allowed.   Generically we expect both modes
to have sizeable branching ratios, although phase
space suppression or the size of the stop mixing angle
can cause one or the other of the two modes to dominate.
In region \heavystop\, the decay $\st_1 \rightarrow \char{1}{}b$
accounts for virtually 100\% of the rate, while
$\char{1}{}\rightarrow\ell\snu$ dominates the
chargino decays, so that stops mainly end up
as a $b \ell \snu$ ($=b \ell +$ invisible) final
state.  In some cases,
$\BR(\char{1}{}\rightarrow W^{*}\neut{1})$ can be
significant ({\it i.e.}\ a few percent), though seldom dominant.
This is because we exclude the small corner of
parameter space where $\mchar{1}\sim\msnu$, since
constraints from LEP161 and LEP172 
suggest that $\BR(\char{1}{}\rightarrow W^{*}\neut{1})$
is probably small for $\char{1}{}$.
Region \degenerate\ combines features from the first two
regions:  
we have $\BR(\st_1\rightarrow \neut{1}c)= 100\%$
(as in region \heavychar),  
but the $\char{1}{}$ decays mostly to $\ell\snu$,
(as in region \heavystop). 
In the cases where the dominant $\char{1}{}$ decay is to $\ell\snu$,
we note that most reported limits on $\mchar{1}{}$ do not apply,
both because the $\char{1}{}$ decay is non-standard and because
the $\char{1}{}$ cross section is reduced by $\snu$
exchange for our range of~$\msnu$.


\section{LEP Cross Sections and Signatures} \label{LEP}

We discuss those signals which are large enough
to be detectable at LEP with $\sim 100$~pb$^{-1}$
per detector at a center of mass energy $\sqrt{s} = 184 \GeV$.  
The cross section
estimates presented below were obtained using the 
{\tt SPYTHIA}\ Monte Carlo~\cite{PYTHIA,SPYTHIA}.

Although $\neut{1}\neut{3}$ and $\snu\bar{\snu}$ production
are among the largest cross sections, they are almost
entirely invisible, as described above.  However, on occasion
we have
$\snu \rightarrow \neut{2}(\rightarrow\neut{1}\gamma)\nu$,
implying $e^{+}e^{-}\rightarrow \snu\bar{\snu}\rightarrow \gamma I$, 
where we use $I$ to stand for a set of invisible particles.  
Since $I$ includes two $\neut{1}$'s and a neutrino,
and one of the $\neut{1}$'s must combine with the neutrino
to form an on-shell sneutrino,
the {\it minimum}\ missing invariant
mass is $\msnu+\mneut{1}\approx 120 \GeV$.
Similarly, on occasion $\neut{1}\neut{3}$ will give rise to
$\neut{1}\snu(\rightarrow\neut{1}\gamma\nu)\bar{\nu}$, with a
missing invariant mass of at least $2\mneut{1}$.
The production of $\neut{2}\neut{3}$ followed by the dominant
$\neut{2}$ and $\neut{3}$ decays also leads to a single photon
plus missing energy:  in fact, this mode accounts for the
majority of the $\gamma I$ total rate.
Photons can also be radiated from the initial electrons
(or from the $t$-channel chargino in $\snu\bar{\snu}$ 
production, although this contribution is small);
the threshold recoil mass here for $\snu\bar{\snu}$ is
about 150 GeV, and for $\neut{1}\neut{3}$ about 120 GeV.
Other channels that can give a photon include
$e^{+}e^{-} \rightarrow \neut{1}\neut{1}$ with a
radiated $\gamma$,
$e^{+}e^{-} \rightarrow \neut{1}\neut{2}(\rightarrow\neut{1}\gamma)$,
etc., but these give a smaller contribution\cite{SUSYRad}.  The 
entire effect can be large, giving an excess over 
the \SM\ single photon
rate for large missing invariant mass as one signature
for supersymmetry.   In the models we are considering, the
total SUSY-related $\gamma I$ rate is typically between 100 and 300 fb.
Once the signal is detected, 
it constrains the $\neut{1}$, $\neut{2}$, 
$\neut{3}$, and $\snu$ masses.

Of course, there is a background for this $\gamma I$ channel
from $\gamma Z(\rightarrow \nu\bar\nu)$
and direct $\gamma\nu\bar\nu$ production via $W$-exchange.  
Most of the background is
not in the region of interest here (missing mass well above $M_Z$),
but enough is that some study is required to determine the best
cuts on the observed photon.
The authors of Ref.~\cite{Gunion} have studied
largely invisible SUSY signatures and how they  might
appear in the $\gamma I$ channel.  The SUSY cases
they examine do not overlap ours, but some of the 
phenomenology is the same.  Their comments on \SM\ backgrounds
are relevant, particularly for $e^{+}e^{-}\gamma$ events
where neither lepton is detected.

Another channel that can give a signal is\footnote{
$\neut{2}\neut{3}$ production
followed by $\neut{2}\rightarrow\neut{1}\gamma$ and 
$\neut{3}\rightarrow\gamma\neut{1}\nu\bar\nu$
provides another (but much smaller) source of $\gamma\gamma I$
events.}
\beq
e^{+}e^{-}\rightarrow\neut{2}\neut{2}
\rightarrow\gamma\gamma I.
\label{TwoPhoton}
\eeq
This channel is particularly interesting because about six
events with missing invariant masses greater than 100 GeV
have been reported\cite{LEPhints} from LEP161 and LEP172 
running, combining all four detectors and both energies.
The events are precisely in the region where we expect a signal
in our models
({\it i.e.}\ missing invariant mass above 100 GeV).

It is very important to know the \SM\ background well, since
the signal may only be a few times larger.  The best estimate of
the background is by S. Ambrosanio~\cite{Sandro}, who has
done a careful calculation at tree-level
using {\tt CompHEP3.0}~\cite{COMPHEP}.  
For photons satisfying the requirements
$E_\gamma > 8\GeV$,
$\vert\cos\theta_\gamma\vert < 0.95$,
and $\mmiss > M_Z + 4\Gamma_Z$, he finds 
a cross section of $20 \pm 2$ fb, which
gives a background of 1.6 events for perfect photon detection
efficiency.  
Assuming an average photon efficiency of 0.8,
the final expected background is about 1.3 events.  We have
made checks using {\tt PYTHIA}~\cite{PYTHIA} (which is
not ideal for such calculations), and find numbers consistent
with Ambrosanio's, and certainly not noticeably larger.
If a signal is found, a precise evaluation of the higher-order
corrections to this background would be useful.  As far as we know,
this has not yet been carried out.

The signal has a missing invariant mass $\mmiss$ larger
than $2\mneut{1}$, and photons
that are never very soft (because we require 
$\mneut{2}-\mneut{1}>20 \GeV$).
The energy range for the photons produced 
in~(\ref{TwoPhoton}) is
\beq
E_{min,max} =
{ {\sqrt{s}} \over {4\mneut{2}^2} }
(\mneut{2}^2 - \mneut{1}^2)
\biggl(1 \mp \sqrt{1-4\mneut{2}^2/s}\thinspace\biggr)
\label{Eminmax}
\eeq
where $\mneut{1}$ and $\mneut{2}$ are the neutralino
masses and $\sqrt{s}$ is the beam energy.  Although
initial state radiation and detector resolution effects
will result in some photons having less energy than this
minimum, the majority of the signal events in our
models will have $E_\gamma > 8 \GeV$.  
In contrast, the background
(from $\gamma\gamma Z(\rightarrow \nu\bar\nu)$ and
$\gamma\gamma\nu\bar\nu$) has a missing invariant mass
distribution which is
concentrated around the $Z$ peak; those background
events which do have large $\mmiss$ tend to contain low-energy 
photons.  
Thus, the analysis should impose a minimum
energy requirement of about 8 GeV for each photon, 
and define three regions for
$\mmiss$, say $\mmiss < M_Z - 10 \GeV$,
$M_Z - 10 \GeV < \mmiss < M_Z + 10 \GeV$, and 
$\mmiss > M_Z + 10 \GeV$.  
The signal we predict is entirely in the
region $\mmiss > M_Z + 10 \GeV$, while little
background is in this region (see above).
Of course, if softer photons or events
from the $Z$ peak are included in the signal region it will
be harder to detect a signal.  Even with our tightly
constrained parameter space there is a
large variation in $\sigma(\gamma\gamma I)$, but the
majority of the parameter space gives cross sections at LEP184
from 50--400 fb.  (If we consider only those models which
imply 3--10 events of $\gamma\gamma I$ at LEP161+LEP172,
then this range narrows to 100--220 fb.)
Note that this channel is independent
of the single photon channel as a SUSY signal.  Once
the signal is detected, it constrains the $\neut{1}$ and
$\neut{2}$ masses.

If the value of $\BR(\neut{2}\rightarrow\neut{1}\gamma)$
is only somewhat larger than 50\%, the process
$e^{+}e^{-}\rightarrow\neut{2}\neut{2}
\rightarrow\ell^{+}\ell^{-}\neut{1}\gamma\neut{1}$
can become important.\footnote{$\neut{1}\neut{4}$ production
also provides a source of $\ell^{+}\ell^{-}\gamma\Emiss$
events (with differing kinematics).  However,
at LEP184 the cross section times branching ratio for
this mode is at most about 10 fb.}
The signal in this case is large
missing energy ($\Emiss > 2\mneut{1}$) and two leptons 
with a  pair mass many widths below the $Z$
($M_{\ell\bar\ell}^2 \alt \mneut{2}^2 - \mneut{1}^2$).
Our parameter space contains models with up to 200 fb in this channel.

\begin{figure}[h]
\vspace*{15cm}
\includegraphics{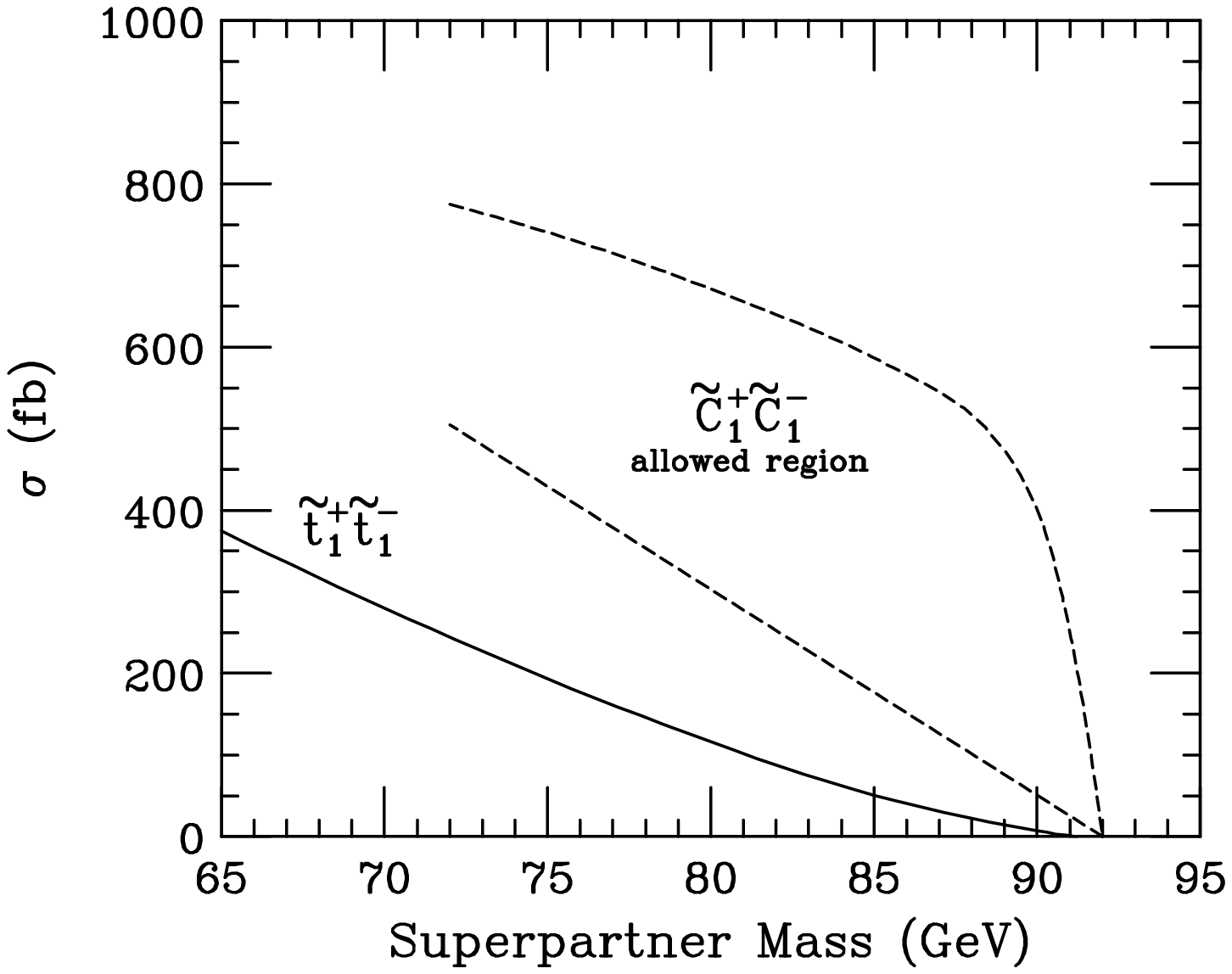}
\vspace{2.0cm}

\caption[]{Estimated total production cross sections for
stops and charginos at LEP ($\sqrt{s} = 184 \GeV$) as a function
of their mass.  For charginos, the cross section is rather
dependent on the details of the model, 
so we indicate a range of values.  Models
with very light ($\alt 72 \GeV$) charginos that satisfy the
constraints described in this paper are very difficult to
construct.
}
\label{LEPfig}
\end{figure}

The other channels at LEP184 which could have sizeable cross sections 
are $\char{1}{+}\char{1}{-}$ and 
$\st_1\bar{\st}_1$ (see Fig.~\ref{LEPfig}).
For charginos, we just combine the
single chargino results above.
In region \heavychar,  chargino decays to $\st_1 b$ and $\ell\snu$
can be comparable, so there are three different
signatures:  $\ell^{+}\ell^{\prime -}\Emiss$ ($\Emiss > 2\msnu$),
$bc\bar{b}\bar{c}\Emiss$ ($\Emiss > 2\mneut{1}$),
and $\ell^{\pm}bc\Emiss$ ($\Emiss > \mneut{1}{+}\msnu$).
Here and below $\ell$ and $\ell^{\prime}$ are any charged leptons,
but $\ell^{\prime}$ can be different from $\ell$.
In regions \heavystop\ and \degenerate, 
the decay $\char{1}{}\rightarrow\ell\snu$
dominates, so chargino pairs 
primarily give $\ell^{+}\ell^{\prime -}\Emiss$
($\Emiss > 2\msnu$).  
Because we have $\mchar{1}{} > \msnu$, the decay
$\char{1}{}\rightarrow W^{*}\neut{1}$ never dominates.
However, it can lead to
$\ell jj\Emiss$ ($\Emiss > \msnu{+}\mneut{1}$) 
as an additional signature, but at a much reduced rate.
In all cases, $\ell^{+}\ell^{\prime -}\Emiss$ ($\Emiss > 2\msnu$)
will be important as the chargino pair signature,
with all combinations of $\ell=e,\mu,\tau$ and 
$\ell^{\prime}=e,\mu,\tau$ possible.  Since
$\snu_e$, $\snu_\mu$, and $\snu_\tau$ are not guaranteed
to be exactly degenerate, the relative number of each type
of lepton pairs cannot be precisely 
predicted.  Once such events are seen,
they will provide information about the sneutrino mass splittings.

For stops we proceed similarly.  
In regions \heavychar\ and \degenerate\ stop pairs 
produce $c\bar{c}\Emiss$ ($\Emiss > 2\mneut{1}$).
In region \heavystop\ stop pairs give the signature
$\ell^{+}\ell^{\prime -} b \bar{b}\Emiss$ ($\Emiss > 2\msnu$).
Note that the $b$ and $\bar{b}$ can be very soft. 

The observation of $\ell^{+}\ell^{\prime -}\Emiss$ ($\Emiss > 2\msnu$)
could signal either charginos or stops.  The presence of a
soft $b\bar{b}$ pair, which might simply appear as a large
hadron multiplicity, would tell us that $\mstop{1} > \mchar{1}$.

The biggest potential background to the $\ell^{+}\ell^{\prime -}\Emiss$
channel is 
$e^{+} e^{-} \rightarrow W^{+} W^{-} 
\rightarrow \ell^{+}\nu_\ell\ell^{-}\bar\nu_\ell$.  However, 
leptons coming from $W$-pair production are fairly stiff
($E^\ell_{min} \approx 24 \GeV$ 
at $\sqrt{s} = 184 \GeV$).\footnote{$WW$ events containing two
leptonic tau decays produce leptons which are
much softer.  However, this mode is suppressed by the
branching ratio factor 
$[\BR(W\rightarrow\tau\nu)\BR(\tau\rightarrow\ell\nu\bar\nu)]^2 =
1.5 \times 10^{-3}$.}
On the other hand, the $\char{1}{}$-$\snu$ mass difference in
our models tends to be smaller than 10 GeV, and is frequently
only a few GeV.  Consequently, the maximum lepton energy coming
from $\char{1}{}$ decay in $\char{1}{+}\char{1}{-}$ events
at LEP184 is only a few GeV.  Thus, a good event selection strategy
would veto energetic leptons, but retain leptons which are as soft
as possible.  

Finally, in some of our models, left-handed
slepton pairs are (barely) light enough to be produced
at LEP184, giving 
$\ell^{+}\ell^{-}\gamma\gamma\Emiss\enspace (\Emiss > 2\mneut{1})$.
Unfortunately, the 
cross section is phase-space limited, and amounts to only
2--3 fb per slepton flavor.


\section{Fermilab} \label{FNAL}

In the scenario described in this paper, many light superpartners 
are produced at Fermilab.  However,
the combination of the near invisibility of $\neut{3}$, $\neut{4}$,
and $\snu$ with the backgrounds at the Tevatron makes for few
good signals.  
The sparticles which give potentially visible
decays are the charginos (especially $\char{2}{}$), $\neut{2}$,
and the lighter stop (and, possibly, the gluino and heavier
squarks:  see the end of this section).  
At this point, some model dependence
enters the discussion, as the relative masses of the $\char{2}{}$
and $\st_1$ are important in determining their decay modes.
However, two signals which are likely to be important are
the inclusive $\gamma\gamma\ETmiss + X$ and $\gamma b \ETmiss + X$
rates.  The first of these two signals is of interest in a 
wider category of models than considered here~\cite{gravitinos},
while the second is special in that it
has no significant parton-level \SM\ background.   

We begin our discussion with $\gamma\gamma\ETmiss + X$.
Both the
Collider Detector at Fermilab (CDF) and D-Zero (D0) collaborations
have reported results on searches for such a 
signal~\cite{GammaGammaData1,GammaGammaData2}.
The CDF results are still preliminary, and do not yet include 
an upper limit on the $\gamma\gamma\ETmiss + X$ rate. 
D0, however, reports the 95\% C.L. upper limit
$\sigma\cdot\BR(p\bar{p} \rightarrow \gamma\gamma\ETmiss + X) < 185$ 
fb for photons satisfying the following cuts:
transverse energy $E_T^\gamma > 12 \GeV$,
pseudorapidity $\vert \eta^\gamma \vert < 1.1,$ and 
missing transverse energy $\ETmiss > 25 \GeV.$
Within the group of models we are studying here, we obtain
this signal from
$\char{2}{+}\char{2}{-}$, $\char{2}{\pm}\neut{2}$,
$\neut{2}\neut{2}$, $\slp_L^{+}\slp_L^{-}$,
and $\slp_R^{+}\slp_R^{-}$ production.
Since the
production cross section for $\char{2}{}$ pairs at the
Tevatron  is large ($\sim 300$--700 fb in our models),
it  potentially provides the largest contribution to
the signal.
First, let us assume that the stop is heavy ($\mstop{1} > \mchar{2}$).
Then,  $\BR(\char{2}{}\rightarrow{\slp}_L\nu) \sim {1\over2}$,
and, including the remaining branching ratios, we obtain a
contribution of around 100 fb.  However, when we account for the
effects of the cuts employed by D0, we find that the total
$\gamma\gamma\ETmiss + X$ rate is never more than 60 fb,
even when the other initial states are added.
Thus, our models are consistent with current Fermilab
search limits.

The expected number of $\gamma\gamma\ETmiss + X$ events actually
decreases if the stop mass is lowered.  First, a smaller
stop mass implies a smaller $\neut{2} \rightarrow \neut{1}\gamma$
branching ratio\cite{SandroLong}.  Second, if $\mstop{1} < \mchar{2}{}$,
then the decay $\char{2}{} \rightarrow \st_1 b$ opens up, allowing
for a $\gamma b \ETmiss + X$ final state when the two charginos
decay differently.  This possibility is especially interesting
since there is no significant
parton-level source of such events within the
Standard Model.  Depending upon the value of 
$\BR(\char{2}{}\rightarrow\st_1 b)$, we estimate that 
the size of the $\gamma b \ETmiss + X$ signal 
from $\char{2}{+}\char{2}{-}$ could be up to 100 fb.  

If $\mstop{1}$ is even somewhat lighter still, the decay 
$t \rightarrow \neut{2} \st_1$ becomes allowed.  In this case,
$t\bar{t}$ production followed by 
$t \rightarrow \neut{2}(\rightarrow \neut{1}\gamma)$
and $\bar{t}\rightarrow W^{-} \bar{b}$ provides an additional
source of $\gamma b \ETmiss + X$ events.  Although 
$\BR(t \rightarrow \neut{2}\st_1)$ tends to be only a few percent
at most, the $t \bar{t}$ production cross section is enormous
($\sim 7$ pb).  Thus, even a modest 1\% value for this branching
ratio can lead to an additional 100 fb of $\gamma b \ETmiss + X$
production.  In connection with this possibility, we note that  
if $t \rightarrow \neut{2} \st_1$ is allowed, then it is likely
(but not certain) that $\mstop{1} < \mchar{1} + m_b$, in which
case the signal becomes $\gamma b c \ETmiss + X$.
Additional consequences for the Tevatron which arise in models 
which allow top-to-stop
decays are discussed in Refs.~\cite{XtraTops,SUwgt}.

Finally, we remark that
if charginos and sleptons are in the mass ranges of Table~\ref{Decays},
then in many models gluinos and squarks of the first two families 
fall in the mass range 200--300 GeV.  These have large cross sections
at Fermilab and might also be observable\cite{XtraTops}, possibly even 
in the present data sample.  The signatures of some such
events could cause them to be included in the top quark sample.


\section{Comments} \label{COMMENTS}

This paper reports the predictions for LEP of a particular,
interesting, region of the SUSY parameter space.  It is worth reporting
these predictions because they are quite different from those
of most SUSY analyses.  For example, the largest cross sections
($\neut{1}\neut{3}$ and $\snu\bar{\snu}$) are almost
completely invisible; they only show up occasionally as a single
photon plus large missing energy.  The cleanest SUSY signature
may be $\gamma\gamma$ plus large missing energy, from 
$\neut{2}\neut{2}\rightarrow\neut{1}\neut{1}\gamma\gamma$.
Chargino pairs mainly give $\ell^{\pm}\ell^{\prime\mp}$ plus
large missing energy, where $\ell$ and $\ell^{\prime}$ can be
different leptons and are soft.  Stop pairs and selectron
pairs are also possible.  If no signals are observed at
LEP184, this SUSY scenario is almost, but not 
quite, eliminated.\footnote{The interpretation
of the large missing mass $\gamma\gamma\Emiss$ events from LEP161 
and LEP172 running as $\neut{2}\neut{2}$ production
would become untenable in the absence
of a signal at LEP184.}
At LEP194, this scenario is completely excluded if no signatures 
are seen with $\sim 100$ pb$^{-1}$ per detector.
Note that most, but not all, sets of masses consistent
with Eqs.~(\ref{Inputs}) and~(\ref{MassH}) are also consistent
with all present data.  This is because if~(\ref{Inputs})
and~(\ref{MassH}) do indeed describe the real world,
then the sparticles are on the verge of being detected.

If such signals are seen, it will be easy (and fun) to extract
from even limited data the remaining parameters of the 
chargino, neutralino and left-handed slepton sectors, $\tan\beta$,
and (if present) light stop to good accuracy, even in the 
most general framework of a softly broken supersymmetric theory.
If a model such as those examined here was indeed
observed, the implications for the structure of the
fundamental theory will be unusually interesting, since
some features are likely to be different from those in 
most minimal SUSY models.  For example, $M_1/M_2$ seems
to be nearer to unity than to ${5\over3}\tan^2\theta_W$, 
and $\mslp{L} < \mslp{R}$ (which could happen, for
example, from the $D$-terms  coming from an 
extra $U(1)$~\cite{KoldaMartin}).  
If these signals are seen it is very likely that the lightest
SUSY Higgs boson is accessible at LEP, and it is
certainly detectable at Fermilab within
a few years.


\acknowledgements

This research is supported in part by the U.S. Department of Energy.
We appreciate conversations with J. Gunion, D. Stickland, D. Treille, 
and G. Wilson.  
We are very grateful to G. Kribs for checking some results and for 
helpful discussions, and to S. Ambrosanio for discussions and
the use of some of his computer programs to check some of the results.
Finally, we thank S. Ambrosanio, G. Kribs, and S. Martin for
reading the manuscript and providing useful feedback.



\vspace*{1cm}

\begin{table}
\caption{Masses and main decays of sparticles relevant to LEP,
for models satisfying the mass hierarchy of Eq.~(\ref{MassH}).
Unless specific quark or lepton flavors are indicated, we 
quote the sum over all flavors.
\label{Decays}}
\begin{tabular}{cccccccc}
&& Mass (GeV)  & \multicolumn{2}{c}{Main Decays} & 
                 \multicolumn{2}{c}{Other Significant Decays} & \\[-0.4cm]
&&   & Mode(s) & Fraction & Mode(s) & Fraction & \\
\tableline
& $\neut{1}$ & 45--65 & stable  & ------ & ------ & ------ & \\
& $\neut{2}$ & 65--85 
             & $\neut{1}\gamma$  
             & 50--85\% 
             & $\neut{1}\ell^{+}\ell^{-}$, $\neut{1}q\bar{q}$ 
             & each $\alt 25\%$ & \\
& $\neut{3}$ & 90--110 
             & $\snu\nu$ 
             & $\agt 93\%$ 
             & $\slp_L^{\pm}\ell^{\mp}$
             & $\alt 7\%$ & \\
& $\neut{4}$ & 115--140 
             & $\snu\nu$ 
             & 75\%--85\% 
             & $\slp_L^{\pm}\ell^{\mp}$
             & $\alt 25\%$ & \\
& $\char{1}{}$ & 75--95 
               & $\biggl\{
                  {\mathstrut\hbox{$\snu\ell\ts$\tablenotemark[1]}\atop
                  \mathstrut\hbox{$\snu\ell$, 
                                  $\st_1 b\ts$\tablenotemark[2]}}$
               &  $\agt 90\%$
               & $\neut{1}\ell\nu$, $\neut{1}q\bar{q}^{\prime}$ 
               & $\alt 10\%$\tablenotemark[3]  &\\[0.15cm]
& $\char{2}{}$ & 110--140 
               & $\biggl\{
                  {\mathstrut\hbox{$\snu\ell$, 
                                  $\slp_L\nu\ts$\tablenotemark[4] }\atop
                 \mathstrut\hbox{$\snu\ell$, $\slp_L\nu$, 
                                 $\st_1 b\ts$\tablenotemark[5]}}$
               &  $\agt 95\%$             
               & $\neut{1}\ell\nu$, $\neut{1}q\bar{q}^{\prime}$ 
               & $\alt 5\%$  &\\[0.1cm]
& $\snu$    & 75--90 
            &  $\neut{1}\nu$  &  $\agt 98\%$ 
            &  $\neut{2}\nu$  &  $\alt 2\%$ \tablenotemark[6]  & \\
& $\slp_L$  & 90--105 
            & $\neut{2}\ell$ & $\agt 94\% $ 
            & $\neut{1}\ell$ & $\alt 6\% $ & \\
& $\st_1$   & 65--115\ts\tablenotemark[7] 
            & $\biggl\{
              {\mathstrut\hbox{$\neut{1}c\ts$\tablenotemark[8]}\atop
              \mathstrut\hbox{$\char{1}{}b\ts$\tablenotemark[9]}}$
            & 100\% 
            & ------ & ------ & \\[0.1cm]
& $h^0$  &  65--100\ts\tablenotemark[10] 
         & $b\bar{b}$ & $\sim 80\%$   
         & $\tau\bar\tau$ & $\sim 9\%$  & \\
\end{tabular}
\tablenotetext[1] {If $\mchar{1}{}<\mstop{1}+m_b$.}
\tablenotetext[2] {If $\mchar{1}{}>\mstop{1}+m_b$.}
\tablenotetext[3] {Phase space suppression of the 2-body decay
modes can enlarge these 3-body modes.}
\tablenotetext[4] {If $\mchar{2}{}<\mstop{1}+m_b$.}
\tablenotetext[5] {If $\mchar{2}{}>\mstop{1}+m_b$.}
\tablenotetext[6] {Violation of this bound requires careful
tuning of the input parameters.}
\tablenotetext[7] {The stop mass could be larger than 115 GeV
without changing the essential features found
in this class of models.}
\tablenotetext[8] {If $\mstop{1}<\mchar{1}+m_b$.}
\tablenotetext[9] {If $\mstop{1}>\mchar{1}+m_b$.}
\tablenotetext[10] {This range is implied by the parameters
examined in this paper but
does not affect the essential features of the other sparticles
in this class of models.}
\end{table}


\end{document}